1

# Experimental assignment of many-electron excitations in the photo-ionization of NiO

J.C. Woicik,[1] J.M. Ablett,[2] N.F. Quackenbush,[1] A.K. Rumaiz,[3] C. Weiland,[1]
T.C. Droubay,[4] and S.A. Chambers[4]

[1]Materials Measurement Science Division, Material Measurement Laboratory, National Institute of Standards and Technology, Gaithersburg, MD 20899, USA
[2]Synchrotron SOLEIL, L'Orme des Merisiers, Saint-Aubin, BP 48, 91192 Gif-sur-Yvette Cedex, France
[3]National Synchrotron Light Source II, Brookhaven National Laboratory, Upton, NY 11973, USA
[4]Physical and Computational Sciences Directorate, Pacific Northwestern National Laboratory, Richland, Washington, 99352, USA

The absorption of a photon and the emission of an electron is not a simple, two-particle process. The complicated many-electron features observed during core photo-ionization can therefore reveal many of the hidden secrets about the ground and excited-state electronic structures of a material. Careful analysis of the photon-energy dependence of the Ni *KLL* Auger de-excitation spectra at and above the Ni 1*s* photo-ionization threshold has identified the satellite structure that appears in both the photo-electron emission and the x-ray absorption spectra of NiO as Ni metal 3*d* $e_g$ → Ni metal 3*d* $e_g$ and O ligand 2*p* $e_g$ → Ni metal 3*d* $e_g$ charge-transfer excitations, respectively. These assignments elucidate the conflicting theoretical predictions of the last five decades in addition to other anomalous effects in the spectroscopy of this unique material.

NiO crystallizes in the cubic rock-salt structure with local octahedral ($O_h$) symmetry. Below its Néel temperature of 525 K, it is an insulating anti-ferromagnet [1], with 4 eV band gap and 1 eV $t_{2g}$-$e_g$ crystal-field splitting for its Ni$^{2+}$ 3$d^8$ $t_{2g\uparrow}^3 t_{2g\downarrow}^3 e_{g\uparrow}^2$ high-spin, Hund's-rule ground state as determined by optical absorption [2]. Its two Ni 3*d* $e_g$ electrons are strongly coupled to their next-nearest neighbors, while its six Ni 3*d* $t_{2g}$ electrons exhibit quasi-core behavior based on its narrow 0.3 eV Ni 3*d* bandwidth [3] and the similarity between its core and valence photo-electron spectra [4]. Simple band-theory arguments suggest that its partially filled metal 3*d* band would make NiO a conductor [5], leading Mott [6] and Hubbard [7] to re-examine the role of electron correlations in narrow-band materials. It is therefore no surprise that despite considerable effort over the last five decades the photo-electron spectra of NiO has remained controversial [8]: The primary or "main" photo-emission line was assigned to direct Ni photo-ionization and its "satellite" to monopole ligand-to-metal charge transfer [9] while a configuration-interaction cluster model found the opposite [10]. To date there has been no experiment put forward that uniquely unravels the hidden physics behind these transitions, although NiO has been studied extensively both experimentally and theoretically by numerous spectroscopies [11-36].

Here we utilize high-energy, resonant photoelectron spectroscopy to *experimentally* identify the nature of the satellite structure that appears in both the photo-electron emission and the x-ray absorption spectra of NiO. Our method is based on the ansatz given by Hedin [37], *"For photon energies which barely are large enough to take the electron above the Fermi level there is clearly no energy available to make satellites (or line shape asymmetry)."* Coupled with Siegbahn's original discovery that the Auger de-excitation spectrum of a core hole retains information of an atom's initial charge state [38], the changes that occur in the Auger line-shape as a function of photon energy at and above a core-ionization threshold can uniquely identify the nature of these many-electron processes.

Figure 1 shows the Ni 1*s* photo-electron spectrum from a 200 Å NiO film grown on a Ag(001) substrate. Also shown are Ni-2*p* core and Ni-3*d* valence spectra [39]. The experiment was performed at the Galaxies beamline of Synchrotron SOLEIL using the high-resolution Si(333) reflection from a Si(111) double-crystal monochromator and a hemispherical electron analyzer the cone of which is oriented parallel to the polarization vector of the incident x-ray beam. Details of the beamline [40] and sample-growth technique [41] have been given previously. We chose to study a NiO film grown on a metallic substrate for our Ni 1*s* measurements to avoid possible charging effects that would be likely due to the insulating nature of NiO and the large amount of secondary-electron emission that is produced during the de-excitation of a Ni 1*s* core hole. Each spectrum presented in Figure 1 is consistent with spectra published in the literature, and they identify the satellite structure that will be discussed. Following early assignment [4], we designate peak *A* as the "main line" and peaks *B* and *C* as the "satellite" loss features in order of their relative binding energies. The main line and the satellite structures of the 2*p* core level are mirrored between its spin-orbit split 2$p_{3/2}$ and 2$p_{1/2}$ components, with the 2$p_{1/2}$ component (not shown) exhibiting greater breadth due to Coster-Kronig decay [42]. Additionally, the higher-energy resolution afforded for the Ni 3*d* data on account of its reduced Lorentzian width indicates that the higher binding-energy satellite *C* is split, by an amount equal to the energy separation of peaks *A* and *B*, suggesting that satellite *C* has contributions arising from both peaks *A* and *B*. Fitting the Ni 1*s* photo-electron spectrum with three components and a Shirley background [43] determines the binding energies of the Ni 1*s* satellites relative to the Ni 1*s* main line to be 1.7 eV and 7.2 eV, respectively. From Figure 1, it is also apparent that the satellite binding energies depend on the *l* value of angular momentum probed.

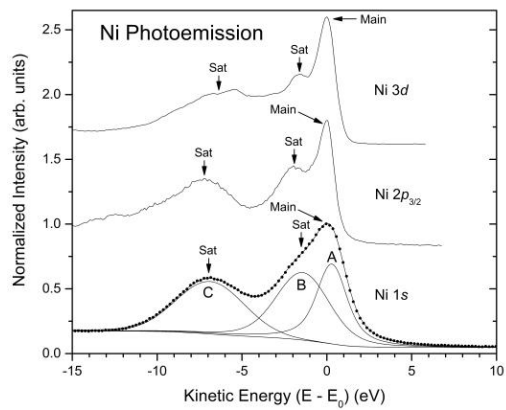

Fig. 1. Ni 1*s*, 2*p*, and 3*d* (valence) photo-electron spectra of NiO. The photon energies were hν = 10,000 eV for the Ni 1*s* level and hν = 2,570 eV for the Ni 2*p* and 3*d* levels, respectively. Also shown is a fit to the data points of the Ni 1*s* spectrum. The spectra have been aligned relative to their maximum intensity

Figure 2 shows the Ni $1s$ x-ray absorption spectra from the NiO film. The data (see inset) were recorded with the synchrotron-beam polarization vector $\varepsilon$ parallel to the NiO [100] and [101] directions while the synchrotron-beam wave-vector $q$ was perpendicular to the NiO [010] direction. The data were recorded by monitoring the Ni $KLL$ Auger partial-electron yield and normalizing to the incident flux taken as the photo-yield from a titanium foil upstream of the sample. For cubic materials, dipole transitions are invariant with respect to $q$ and $\varepsilon$ [44]; consequently, the feature at 8332.5 eV has been identified as a Ni $1s \rightarrow 3d$ quadrupolar transition [45], and more specifically as a Ni $1s \rightarrow 3d\ e_g$ transition, as the $e_g$ orbitals have their maximum electron density oriented along the crystallographic [100], [010], and [001] directions, and the quadrupolar selection rules would therefore either minimize ($\varepsilon$ // [100]) or maximize ($\varepsilon$ // [101]) their intensity for these orientations. The Ni $1s \rightarrow 3d\ t_{2g}$ transition is not observed due to the crystal-field split $t_{2g}^6 e_g^2$ high-spin, Hund's-rule ground state of the Ni$^{2+}$ $d^8$ ion [46]: Each triply degenerate $t_{2g}$ orbital ($d_{xy}$, $d_{yz}$, and $d_{zx}$) is occupied by two electrons, and each doubly degenerate $e_g$ orbital ($d_{3z^2-r^2}$ and $d_{x^2-y^2}$) is occupied by one electron, both being either spin up or spin down. This high-spin configuration is consistent with the NiO anti-ferromagnetic ground state and the Ni$^{2+}$ moment of 2 Bohr magnetons [47]. The Ni $1s \rightarrow 3d$ transition appears sharp, rather than band-like, due to the excitonic attraction between the Ni $1s$ core hole and the electron in the Ni $3d$ level.

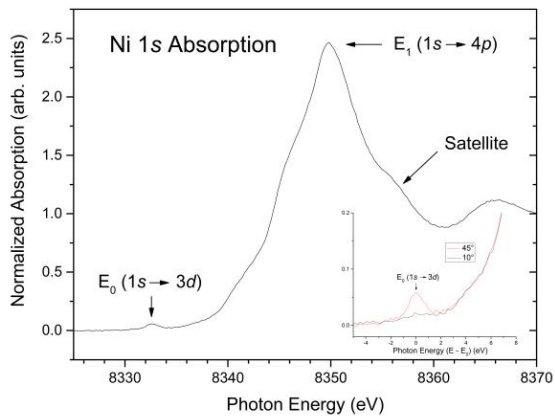

Fig. 2. Ni $1s$ x-ray absorption spectrum of NiO. The inset shows the polarization dependence of the Ni $1s \rightarrow 3d$ quadrupolar transition. Also indicated are the energies of the $1s \rightarrow 4p$ dipole transition and the many-electron satellite.

Figure 3 shows the Ni $KLL$ Auger de-excitation spectra for photon energies around the Ni $1s \rightarrow 3d$ transition. For the Auger measurements, the sample was oriented at 45° x-ray incidence; i.e., ($\varepsilon$ // [101]). Note the distinct multiplet structure that arises from the two holes left in the Ni $2p$ level following $KLL$ decay (see inset) [46]: $^1S_0$ ($K$-$L_2L_2$), $^1D_2$ ($K$-$L_2L_3$), and $^3P_2$ ($K$-$L_3L_3$). These term splittings agree with theoretical calculations [48] and experimental observations [49] for Ni metal; however, for NiO each term is shifted by approximately 6 eV relative to its value in the metal due to the chemical bonding between nickel and oxygen [38]. Also apparent is the large Auger satellite that occurs for each configuration at approximately 9 eV loss. This large satellite is not observed in Ni metal, suggesting that it has a similar electronic origin to that of peak $C$ in the photo-electron spectra [50]. Equally important is the distinctive Auger resonant-Raman shift [51,52] of the main $^1D_2$ line and its satellite with photon energy around the Ni $1s \rightarrow 3d$ transition that confirms the localized nature of this transition: At threshold, the Auger peak sharpens, and it disperses linearly with photon energy due to the conservation of energy between the incident photon, the electron in its $3d$ excitonic bound state, and the Auger electron in the vacuum. At its maximum intensity, the center of the primary $^1D_2$ transition occurs at 6553.5 eV, as indicated in the Figure, while its satellite occurs at 6545 eV.

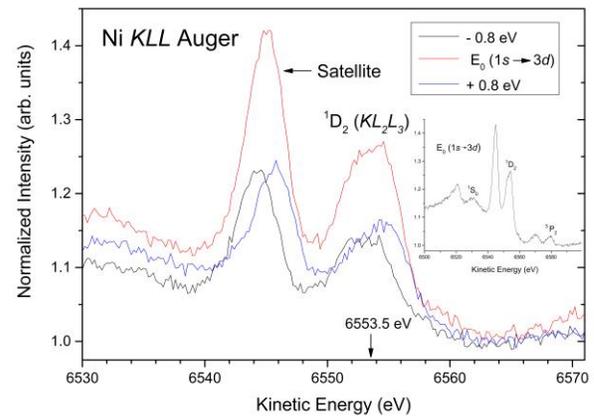

Fig. 3. Photon-energy dependence of the Ni $^1D_2$ $K$-$L_2L_3$ Auger decay of NiO around the photon energy of the Ni $1s \rightarrow 3d$ resonance, as indicated in Figure 2. The inset shows the entire spectrum at resonance.

Note the widths of the main $^1D_2$ line and its satellite. Despite its narrowing at resonance, the main $^1D_2$ line still appears broad and asymmetric, while its satellite is significantly sharper. The splitting of the main line is determined to be 2.1 eV by a two-Gaussian fit to the spectrum. This splitting may be attributed to the additional term splitting of the $2p^4$ Auger final state due to the presence of the single unpaired electron in the Ni $3d\ e_g$ orbital at the photon energy of the Ni $1s$ to $3d$ transition [46], in addition to the splitting of the $^1D_2$ term by the ligand crystal field. As discussed by Cotton [53], the splitting of a D term will be just the same as the splitting of the set of one-electron $d$ orbitals. Taken together, we believe that this is a unique experimental observation for a solid. The fact that the satellite appears narrower than its main line will be addressed further below and shown to be consistent with our experimental assignment of the Ni photo-electron spectra.

Figure 4 shows the $^1D_2$ Auger transition, but now plotted for photon energies equal to the $1s \rightarrow 4p$ transition (the maximum of the Ni $1s$ absorption in Figure 2), between the maximum and the shoulder that occurs approximately 7 eV above it (labeled "Satellite" in Figure 2), at the shoulder itself, and at the trough immediately above the shoulder. It should be emphasized that all of the photon energies studied in Figure 4 are at least 17 eV above the photon energy of the $1s \rightarrow 3d$ transition, and they therefore probe the electron dynamics that occur as the Ni $1s$ electron transits to the continuum as opposed to the resonant behavior that occurs when it is trapped in its $3d$ bound state below it. The shoulder has been identified as a many-electron feature because it does not appear in single-particle calculations of the x-ray absorption coefficient, but it does appear when the single-particle theory is convoluted with the Ni $1s$ photo-electron spectrum [30] (as well as with more sophisticated spectral functions [31]). The fact that this feature appears at a photon energy relative to the maximum of the $1s \rightarrow 4p$ transition that is identical to the binding energy of peak $C$ in the Ni $1s$ photo-electron spectrum also indicates that its origin is the same for both spectra.

The Auger spectra in Figure 4 reveal additional intensity that is shifted by approximately 3 eV to higher kinetic energy (lower binding energy) relative to the main Auger line that turns on at the photon energy of the satellite (indicated as +6.8 eV in the figure) and then reduces in intensity as the photon energy is increased. If this feature were due to an additional intrinsic loss of the primary Auger decay, it would occur at a kinetic energy below rather than above its parent line. Consequently, this feature must be due to a well screened charge-transfer state associated with the $1s \rightarrow 4p$ transition that requires an additional 7 eV of work to create [54]. Threshold phenomena and satellite structure have been observed previously in the Auger spectra of Ni metal for both the *LMM* [55] and *KLL* [49] transitions, but the satellites observed for the metal occur only on the low kinetic-energy (high binding-energy) side of the parent Auger line that identifies them as "shake-off" rather than "shake-on" charge-transfer processes. This observation is also consistent with the resonant Auger spectra of Ar gas for which only shake-off can occur [56].

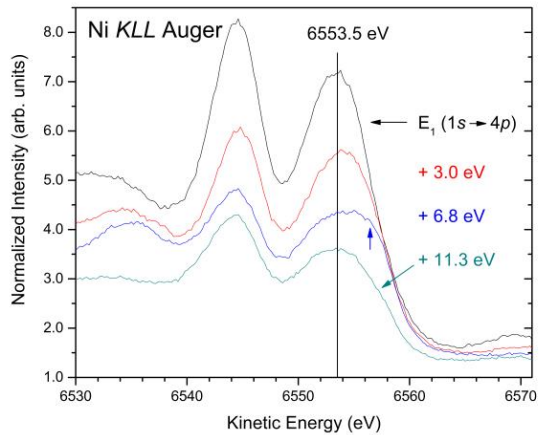

Fig. 4. Photon-energy dependence of the Ni $^1D_2$ $K$-$L_2L_3$ Auger decay of NiO beginning at the maximum of the Ni $1s \rightarrow 4p$ transition and concluding at its trough, as indicated in Figure 2. Note the turn on of the high kinetic-energy intensity in the Auger spectrum at the photon energy of the satellite (indicated as +6.8 eV in the Figure). The vertical arrow marks the additional intensity at +3 eV (see text).

The additional intensity in the NiO Auger spectrum appearing on the high kinetic-energy side of the primary Auger peak with excess photon energy above the $1s \rightarrow 4p$ transition clearly identifies it a shake-on charge-transfer process that screens the $1s$ core hole prior to the Auger decay. The observed approximate 3 eV shift is consistent with a single electron transfer from ligand to metal based on a linear estimate of the Ni *KLL* Auger energy with oxidation state for Ni metal (Ni$^0$: 6559.2 eV [49]) and NiO (Ni$^{+2}$: 6553.5 eV): (6559.2 − 6553.5)/2 = 2.9 eV per electron. As indicated in Figure 6, the Ni atom is actually *not* photo-ionized after either the resonant $1s \rightarrow 3d$ transition or the shake-on charge-transfer processes both of which leave an additional electron in the Ni $3d$ orbital. As the same satellite feature occurs in the photo-electron spectra of Ni doped MgO that has no metal $3d$ electrons [21], we *experimentally* identify this feature as ligand-to-metal charge transfer; i.e., O ligand $2p$ $e_g$ → Ni metal $3d$ $e_g$ monopole charge transfer within the sudden approximation of quantum mechanics [9].

The fact that the Auger-peak energy at the maxima of the $1s \rightarrow 3d$ and the $1s \rightarrow 4p$ transitions occurs at the same kinetic energy would indicate that the Auger electron is emitted with the same amount of core-hole screening for both transitions. However, due to the delocalized band-like nature of the $1s \rightarrow 4p$ transition, this result suggests that an additional charge-transfer process has occurred prior to the maximum of the $1s \rightarrow 4p$ transition. To explore this possibility further, Figure 5 shows the evolution of the $^1D_2$ Auger spectrum, but plotted now as a function of photon energy beginning at the threshold of the $1s \rightarrow 4p$ transition (3.0 eV above the $1s \rightarrow 3d$ transition) and concluding at its maximum. The Auger peak first appears at kinetic energy 1.2 eV below its kinetic-energy maximum. It begins to shift towards higher kinetic energy at photon energy 4.5 eV above the $1s \rightarrow 4p$ threshold (indicated as +7.5 eV in the figure); it then asymptotically approaches its maximum kinetic-energy value that occurs at the maximum of the $1s \rightarrow 4p$ transition. Realizing that the first satellite of the Ni $1s$ photo-electron spectrum peak *B* would turn on within this photon-energy range and realizing also that peak *B* is

*not* observed in the photo-electron spectra of Ni doped MgO that again has no metal $3d$ electrons [21], we *experimentally* assign this feature to Ni metal $3d$ $e_g$ → Ni metal $3d$ $e_g$ transitions arising from a neighboring Ni site. This transition is naturally spin allowed ($\Delta S = 0$) due to the anti-ferromagnetic coupling between next-nearest neighbor Ni atoms, and it also satisfies the monopole selection rules of the sudden approximation [9]. The double-peaked structure of the Ni photo-electron spectra has previously been attributed to nonlocal intra-site screening from neighboring Ni clusters [18], and our data are consistent with this conclusion. We should also add that the Ni $1s$ x-ray absorption edge itself shifts by a full 1 eV when the Ni $1s$ photo-electron main line (peak *A*) is suppressed in the convolution of the single-particle theory [30], supporting the above finding that two distinct absorption processes contribute to the threshold behavior of the Ni $1s \rightarrow 4p$ transition.

It may appear from Figures 4 and 5 that satellite *B* provides less screening than the ligand-to-metal charge transfer assigned for peak *C* because the Auger peak when recorded at the photon energy of satellite *B* occurs at lower kinetic energy than when recorded at the photon energy of satellite *C*. However, we suggest that peak *C* may involve a double-charge transfer (i.e., peak *C* may arise from both peaks *A* and *B*, with the latter contribution already having been screened). This conjecture is consistent with the physics associated with the Zhang-Rice singlet state [57] and the multiple-charge states solution of the Anderson Hamiltonian [58] that binds holes created on neighboring metal atoms by hybridization with their ligands [18].

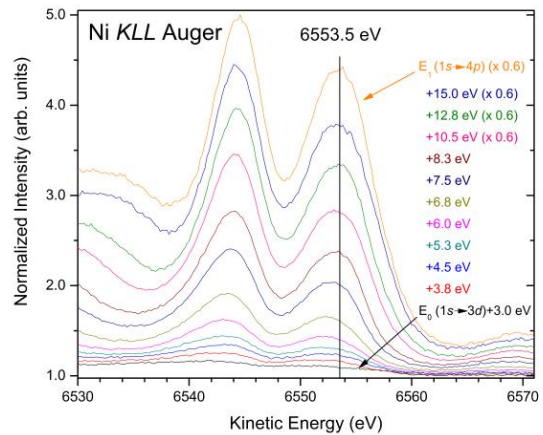

Fig. 5. Photon-energy dependence of the Ni $^1D_2$ $K$-$L_2L_3$ Auger decay of NiO beginning at the threshold of the Ni $1s \rightarrow 4p$ transition and concluding at its maximum, as indicated in Figure 2. Note that shift of the Auger peak by 1.2 eV that begins 4.5 eV above the threshold of the $1s \rightarrow 4p$ transition (indicated as +7.5 eV in the Figure) and concludes at its maximum.

Recently Kas et al. [32] have applied *first-principles*, real-time density-functional theory to the many-body problem of charge-transfer satellites in correlated materials. The calculations reproduce all features of the Ni $1s$ and $2p$ photo-electron spectra seen in Figure 1. The calculations model the core-hole interaction with the Yukawa potential that explicitly neglects exchange interactions, thereby demonstrating that multiplet effects may be considered as detailed perturbations to the satellite structures observed. The calculations also find a splitting of satellite *C* that is consistent with our experimental identification of multiple charge transfer. The interpretation of the calculations, however, assigns the main line *A* to initial charge transfer from ligand to metal reflecting a "well-screened" core hole, while the satellites *B* and *C* reflect charge transfer back to the ligands and a more weakly screened core hole. Had this interpretation been valid, the Auger intensity observed in our spectra at excess photon energy would appear on the low kinetic-energy (high binding-energy) side of the primary Auger line due to the additional Coulomb attraction of the fully ionized or un-screened core-hole state. We should stress that our data in general do not support calculations that attribute the Ni satellite structure to direct photo-ionization.

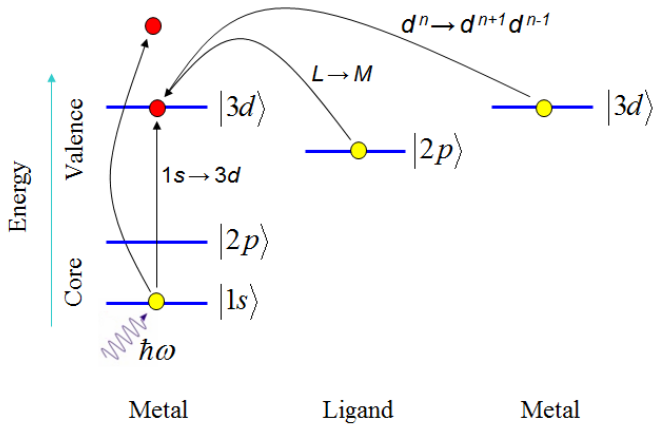

Fig. 6. Different initial states for the Ni *KLL* Auger decay of NiO that are produced during Ni 1*s* x-ray absorption. From left to right: (a) Direct Ni 1*s* photo-ionization. (b) Resonant excitation of the Ni 1*s* electron to the Ni 3*d* level. (c) Ni 1*s* photo-ionization accompanied by $L \to M$ charge transfer. (d) Ni 1*s* photo-ionization accompanied by $M \to M$ ($d^n \to d^{n+1}d^{n-1}$) charge transfer. Note that the latter three processes do not photo-ionize the Ni atom (see text).

We will now address the intriguing physics behind our discovery of an anomalously narrow satellite line. At the $1s \to 3d$ resonance, the final state of the absorption process has an extra electron in the Ni 3*d* level, and this state decays via *KLL* Auger decay: $1s^1 3d^9 \to 2p^4 3d^9 + e^-$. However, for the Auger satellite, charge transfer produces an Auger final state with ten 3*d* electrons: $1s^1 3d^9 \to 2p^4 3d^{10}\underline{L} + e^-$. This now *fully occupied* 3*d* level cannot couple (i.e., there will be no additional multiplet splitting) to the *KLL* Auger terms thereby reducing the width of the satellite relative to the main line. Note as well that the Auger satellite occurs with loss energy 9 eV; i.e., 2 eV greater than what is found in the Ni 1*s* photo-electron spectrum due to the strong repulsion of the additional electron in the Ni 3*d* level. The relative intensity of the satellite is also found to be significantly reduced for Auger transitions that follow charge transfer. The transitions experimentally identified in this work are illustrated in Figure 6.

In conclusion, by measuring the photon-energy dependence of the Ni *KLL* Auger de-excitation spectrum at and above the Ni 1*s* photo-ionization threshold, we have experimentally determined the nature of the satellite structure that appears in both the photo-electron emission and the x-ray absorption spectra of NiO. The amount of core-hole screening present at the satellite binding energies identifies these structures as shake-on charge-transfer excitations that occur in response to the sudden creation of the core hole. We have also demonstrated that charge transfer can produce anomalously narrow satellite lines through its unique ability to fill atomic subshells. This study should therefore help advance *first-principles* methods that predict solid-state electronic structure by providing experimental assignment of this and other photo-ionization spectra.

We acknowledge SOLEIL for provision of synchrotron radiation facilities (Proposal No. 20170393). The work at PNNL was supported by the U.S. Department of Energy, Office of Science, Division of Materials Sciences and Engineering.